\documentstyle[11pt]{article}
\begin{document}

\begin{flushright}
ULB--TH--99/25\\
October 1999
\end{flushright}

\vspace{1.5cm}

\begin{center}
{\LARGE Diffeomorphisms and Weyl tranformations \\in $AdS_3$
gravity}\footnote{
Presented at the TMR European program meeting ``Quantum aspects
of gauge theories, supersymmetry and unification'', ENS, Paris, France,
1-7 September 1999.} 

\vspace{1cm}

{\large Karin Bautier}\footnote{E-mail: kbautier@ulb.ac.be}

\addtocounter{footnote}{-2}

\vspace{.4cm}
{\it Service de Physique Th\'eorique et Math\'ematique}\\
{\it Universit\'e Libre de Bruxelles, Campus Plaine, C.P.231}\\
{\it Boulevard du Triomphe, B-1050 Bruxelles, Belgium}\\

\end{center}

\vspace{2cm}

\begin{abstract}
It is shown that the $AdS_3$ gravity action with boundary terms is non
invariant under diffeomorphisms and that its Lie derivative has the form
of the Weyl anomaly in two dimensions. This variation is compensated by a
Weyl transformation of the boundary metric when the radial derivative of
the metric on the boundary is expressed in terms of the stress tensor of a
Liouville field. The obtained invariance of the action under the 
combined transformation of a diffeomorphism and a Weyl transformation 
allows to interpret the computed Lie derivative as minus the Weyl anomaly
of the two-dimensional effective action. 
\end{abstract}  

\newpage

Brown and Henneaux have shown \cite{brown} that the asymptotic symmetry
group of anti-de Sitter gravity in three dimensions is the conformal group
in two dimensions with a central charge $c=3l/2G$. We are interested in
understanding this central charge as the Weyl anomaly of the
two-dimensional effective action. This anomaly has been calculated in
\cite{skenderis} by means of a regularization procedure and for a constant
Weyl parameter\footnote{In a recent paper, the anomaly has been obtained 
for arbitrary Weyl parameters by computing the boundary effective action 
\cite{skenderis2}.}. Here we will relate it to the non invariance of the
action under diffeomorphisms. We compute the Lie derivative of the
$AdS_3$ on-shell action and show that it has the form of the Weyl anomaly
in two dimensions with the value of the Brown-Henneaux central charge.
However the variation of the action can be compensated by a
Weyl transformation on the boundary, provided that the radial derivative
of the metric on the boundary is expressed in terms of the stress tensor
of a Liouville field. Moreover the Liouville equation and the Einstein
equation are shown to be satisfied at the same time. Therefore the
invariance of the action under the combined transformation of the
diffeomorphism and the Weyl transformation is established and the Lie
derivative of the action can be interpreted as minus the Weyl anomaly of
the two-dimensional effective action. In this way, the relation between
$AdS_3$ gravity and Liouville theory \cite{vandriel} is recovered at the
level of the Weyl transformations properties and of the equations of
motion. The work presented here was done in collaboration with F. Englert,
M. Rooman and P. Spindel, and a extended version of these results will be
published separately \cite{papier}.  

The action considered is the Einstein-Hilbert action for three-dimensional
gravity with a negative cosmological constant, improved by the
Gibbons-Hawking surface term and a constant surface term:
\begin{equation}
S=\frac{1}{16\pi G} \int_M \sqrt{-^{(3)}g}(^{(3)}R+\frac{2}{l^2})d^3x
+\frac{1}{8\pi G} \int_{\partial M}  \sqrt{-^{(3)}g}
(2D_{\mu} n^{\mu}+\frac{1}{l})d^2x.
\end{equation}
The constant term makes the Lie derivative of the action finite. The
Gibbons-Hawking term gets rid of the second derivatives of the metric and
makes the action stationary on the equations of motion for small 
variations of the metric that keep it fixed on the boundary. Here, because
of the cosmological constant in the Einstein equations, the metric
diverges at spatial infinity and the boundary metric is ill-defined. Yet a
conformal class of boundary metrics can be builded by multiplying the
metric by an arbitrary function (which is called the defining function)
that vanishes on the boundary and taking the value of this product on the
boundary as the boundary metric. As this metric changes by a conformal
transformation when the defining function is changed, a conformal
equivalence class of boundary metrics has been defined. 

According to Fefferman and Graham \cite{fefferman}, a suitable choice of
coordinates allows to put any solution of the Einstein equations in the
form:
\begin{equation}
^{(3)}g_{\mu\nu}dx^\mu dx^\nu
=\frac{l^2}{4y^2}dy^2+\frac{1}{y}g_{ij}(y,x)dx^i dx^j,
\label{metric}
\end{equation}
where $y=0$ on the boundary.
The metric $g_{ij}$ appearing in (\ref{metric}) can be expanded order by
order in the radial coordinate $y$, leading to the following
development:
\begin{equation}
^{(3)}g_{\mu\nu}dx^\mu dx^\nu
=\frac{l^2}{4y^2}dy^2+(\frac{1}{y}g_{(0)ij}+g_{(2)ij})dx^i dx^j
+{\cal O}(y),
\label{metric2}
\end{equation}
where $g_{(0)ij}=g_{ij}(y=0,x)$ is a representative of the conformal class
of boundary metrics and $g_{(2)ij}$ is the first radial derivative of
$g_{ij}$ on the boundary. The Einstein equations constraint the metric
$g_{ij}$ order by order in $y$. The equations of motion for $g_{(2)}$ are
given by:  
\begin{eqnarray}
Tr(g_{(0)}^{-1} g_{(2)}) = -\frac{l^2}{2}R(g_{(0)}),& &\label{einstein1}\\ 
D_i(g_{(0)}^{-1ik} g_{(2)kj}) - D_j (g_{(0)}^{-1ik}g_{(2)ik}) &=& 0.
\label{einstein2}
\end{eqnarray}
We see that, unlike in higher dimensions where $g_{(2)}$ is uniquely
determined by the Einstein equations, there is an equation of motion for
its trace-free part. When $g_{(2)}$ is chosen, the rest of the solution
is determined. This shows that the on-shell action depends on the boundary
metric $g_{(0)}$ and on the trace-free part of $g_{(2)}$.

We consider now diffeomorphisms that keep the metric in the form
(\ref{metric}), i.e. such that $\delta_{diff}{}^{(3)}g_{yy}
=\delta_{diff}{}^{(3)}g_{yi}=0$. They are given by: 
\begin{eqnarray}
\delta_{diff}y &=& 2\delta\sigma(x) y, \\
\delta_{diff}x^i &=& -\frac{l^2}{4}\int_0^y g^{ij}(x,y')
\delta\sigma_{,j} dy',
\end{eqnarray}
and induce on $g_{(0)}$ and $g_{(2)}$ the following transformations:
\begin{eqnarray}
\delta_{diff}g_{(0)ij} &=& -2\delta\sigma g_{(0)ij},
\label{diff1} \\
\delta_{diff}g_{(2)ij} &=& -l^2 D_i \partial_j \delta\sigma.
\label{diff2}
\end{eqnarray}
This diffeomorphism keeps $g_{(0)}$ in its conformal class (which is
unique in two dimensions). Yet the Lie derivative of the action on the
equations of motion gives:
\begin{equation}
\delta_{diff}S=\frac{l}{16\pi G} \int_{\partial M}
\sqrt{-g_{(0)}}R(g_{(0)})\delta\sigma d^2x,\label{anomaly}
\end{equation}
showing that the action is not invariant under the diffeomorphism. 

To restore the invariance, we would like to compensate the above
diffeomorphism by a Weyl transformation on the boundary:
\begin{equation}
\delta_W g_{(0)}=2\delta\sigma g_{(0)},
\end{equation} 
which clearly cancels the tranformation (\ref{diff1}) induced by the
diffeomorphism on $g_{(0)}$. We recall that the on-shell action depends
not only on the boundary metric $g_{(0)}$ but also on the trace-free part
of $g_{(2)}$ because there correspond different $g_{(2)}$'s solutions for
the same $g_{(0)}$. During the Weyl transformation, this further degree of
freedom needs to be controlled to ensure that we are back to the initial
solution, i.e. the transformation properties of $g_{(2)}$ must be
specified in order to compensate exactly its variation (\ref{diff2}) under
the diffeomorphism.
To induce this transformation, we note that the dependence of the action 
on the trace-free part of $g_{(2)}$ can be expressed in terms of a field
$\phi$ living on the boundary. We take for $\phi$ the following conformal
weight:
\begin{equation}
\delta_W\phi=-\delta\sigma,
\end{equation}                         
and look for an expression for $g_{(2)}$ as a functional of $g_{(0)}$ and
$\phi$.

We can check that the relation for $g_{(2)}$ in terms of $g_{(0)}$ and
$\phi$ that implies the correct transformation for $g_{(2)}$ under the
Weyl transformation is given by the following expression:
\begin{equation}
g_{(2)ij}=l^2[-D_i\partial_j\phi+\partial_i\phi\partial_j\phi
+g_{(0)ij}(\lambda e^{2\phi}-\frac{1}{2}\partial^k\phi\partial_k\phi)].
\label{phi1}
\end{equation}
It can be written more compactly in terms of the on-shell Liouville stress
tensor $T_{ij}$:
\begin{equation}
\frac{1}{l^2}g_{(2)ij}=\frac{8\pi G}{l}T_{ij}-\frac{1}{2}g_{(0)ij}R,
\label{phi2}
\end{equation}
where $T_{ij}$ is derived from the Liouville action:
\begin{equation}
S=\frac{l}{8\pi G}\int(\frac{1}{2}\sqrt{-g_{(0)}} g_{(0)}^{ij}
\partial_i\phi \partial_j\phi +\frac{1}{2}\sqrt{-g_{(0)}}R\phi +\lambda
\sqrt{-g_{(0)}} e^{2\phi})d^2x.
\end{equation}
The constants in front of the action and $T_{ij}$ have been adjusted to
match the value of the Weyl anomaly that will be computed below.      

Under the Weyl transformations of $g_{(0)}$ and $\phi$,
\begin{eqnarray}
\delta_W g_{(0)} &=& 2\delta\sigma g_{(0)},\label{weyl1}\\
\delta_W\phi &=& -\delta\sigma \label{weyl2},
\end{eqnarray}  
equation (\ref{phi1}) implies the searched-for transformation for
$g_{(2)}$:
\begin{equation}
\delta_W g_{(2)ij}=l^2 D_i \partial_j \delta\sigma.
\end{equation}  
Moreover, we see that the field $\phi$ satisfies the Liouville equation
when the Einstein equation (\ref{einstein1}) on the trace of $g_{(2)}$ is
satisfied:
\begin{equation}
Tr(g_{(0)}^{-1} g_{(2)})+\frac{l^2}{2}R
=l^2(-\Box\phi+\frac{1}{2}R+2\lambda e^{2\phi})=0.
\label{liouville}
\end{equation}
The Einstein equation (\ref{einstein2}) on the trace-free part of
$g_{(2)}$ is then automatically verified:   
\begin{equation}
D_i(g_{(0)}^{-1ik}g_{(2)kj}) - D_j (g_{(0)}^{-1ik}g_{(2)ik})
= l^2 \partial_j\phi(\Box\phi-\frac{1}{2}R-2\lambda e^{2\phi})=0,
\label{conservation}
\end{equation}
and expresses the conservation of the Liouville stress tensor. Indeed it
can be written, with the use of (\ref{liouville}), as:
\begin{equation}
l^2 \frac{8\pi G}{l} D_i T^i_{\ j}=0.
\end{equation}

After doing the diffeomorphism (\ref{diff1},\ref{diff2}) and the Weyl
transformation (\ref{weyl1},\ref{weyl2}), we are back to the initial
solution and the action is invariant under this combined transformation:
\begin{equation}
(\delta_{diff}+\delta_W)S=0.
\end{equation} 
The value of the Weyl anomaly of the two-dimensional effective action can
then be deduced from expression (\ref{anomaly}):
\begin{equation}
\delta_W S=-\delta_{diff}S=-\int_{\partial M}\sqrt{-g_{(0)}}
A\delta\sigma,
\end{equation}
with
\begin{equation}
A=\frac{3l}{2G}\frac{R}{24\pi}.     
\end{equation}
We recover in this way the central charge of the Brown-Henneaux asymptotic
algebra \cite{brown}.   

\subsection*{Acknowledgements}
I would like to thank my collaborators on the work which was the subject
of this talk, F. Englert, M. Rooman and P. Spindel. This work has been
partly supported by the ``Actions de
Recherche Concert{\'e}es" of the ``Direction de la Recherche
Scientifique - Communaut{\'e} Fran{\c c}aise de Belgique" and by
IISN - Belgium (convention 4.4505.86).
The author is ``Chercheur F.R.I.A.'' (Belgium).

\end{document}